%% file: main.tex
\documentclass[runningheads]{llncs}
\usepackage[T1]{fontenc}

\usepackage{listings}
\usepackage{enumitem}
\usepackage{float}
\usepackage{caption}
\usepackage{subcaption}
\usepackage{multirow}
\usepackage[table]{xcolor}
\usepackage{graphicx}

\definecolor{codegreen}{rgb}{0,0.6,0}
\definecolor{codegray}{rgb}{0.5,0.5,0.5}
\definecolor{codepurple}{rgb}{0.58,0,0.82}
\definecolor{backcolour}{rgb}{0.95,0.95,0.92}

\lstdefinestyle{mystyle}{
    commentstyle=\color{codegreen},
    keywordstyle=\color{magenta},
    numberstyle=\tiny\color{codegray},
    stringstyle=\color{codepurple},
    basicstyle=\ttfamily\footnotesize,
    breakatwhitespace=false,         
    breaklines=true,                 
    captionpos=b,                    
    keepspaces=true,                 
    numbers=left,                    
    numbersep=5pt,                  
    showspaces=false,                
    showstringspaces=false,
    showtabs=false,                  
    tabsize=2,
    frame=lines
}

\setlist[itemize]{leftmargin=*}
\setlist[enumerate]{leftmargin=*}

\begin{document}

\title{Contemplating a Lightweight Communication Interface for Asynchronous Many-Task Systems}

\titlerunning{Contemplating a Lightweight Communication Interface for AMTs}
\author{Jiakun Yan\orcidID{0000-0002-6917-5525} \and
Marc Snir\orcidID{0000-0002-3504-2468}}

\authorrunning{J. Yan \and M. Snir}
\institute{University of Illinois Urbana-Champaign, Urbana IL 61801, USA\\
\email{\{jiakuny3,snir\}@illinois.edu}}

\maketitle

\input{sections/abstract}
\input{sections/introduction}
\input{sections/interface}
\input{sections/binding}
\input{sections/performance}
\input{sections/related_work}
\input{sections/conclusion}

\input{output.bbl}
\end{document}

%% file: sections/abstract.tex
\begin{abstract}
Asynchronous Many-Task Systems (AMTs) exhibit different communication patterns from traditional High-Performance Computing (HPC) applications, characterized by asynchrony, concurrency, and multithreading. Existing communication libraries usually do not support AMTs' communication requirements in the most direct and efficient ways. The Lightweight Communication Interface (LCI) is an experimental communication library aiming to push for efficient communication support for AMTs. This paper presents the design for a new LCI C++ interface and its rationale. With a new C++ \emph{objectized flexible functions} idiom, the new interface aims for the following features: (a) a concise but expressive interface for all common point-to-point communication primitives and completion mechanisms, (b) a fine-grained resource mapping scheme for library interoperation, multithreaded performance isolation, and flexibility (c) a set of optional parameters and overridable classes for users to incrementally fine-tune the runtime behavior.

\end{abstract}

%% file: sections/introduction.tex
\section{Introduction}
MPI\cite{mpi10} has been the de-facto standard for parallel computing for more than 30 years. Mainstream MPI applications\cite{tufo1999nek5000,balay2019petsc,thompson2022lammps} typically follow the Bulk-Synchronous Parallel (BSP) model. They feature coarse-grained, collective-style communication, limited communication overlap, and single-threaded communication. Historically, they have driven the MPI community’s optimization efforts. 

Heterogeneous architectures and irregular applications~\cite{hofmeyr2020metahipmer,EarthSystemPaRSEC2024Abdulah,Yadav2023legate_sparse} call for new programming models, and the \emph{Asynchronous Many-Task} (AMT) model emerges as a promising alternative~\cite{kale1993CHARMPortableConcurrent,augonnet2009StarPUUnifiedPlatform,bauer2012LegionExpressingLocality,bosilca2013PaRSECExploitingHeterogeneitya,Kaiser2022HPX}. However, AMTs communicate in a different manner: multiple threads can launch communication concurrently, messages are mostly point-to-point and fine-grained, there can be many pending communication requests simultaneously, and global synchronization points are rare. Unfortunately, many of these characteristics are not well supported by existing communication libraries such as MPI and GASNet-EX~\cite{bonachea_gasnet-ex_2018} and even the lower network portability layer such as UCX~\cite{shamis2015ucx} and libfabric~\cite{libfabric}. 
AMTs must be better supported in this regard if we want to fully exploit AMTs' advantages in overlapping and adaptiveness with a larger number of more fine-grained tasks while scaling to larger node counts. 

The Lightweight Communication Interface (LCI)~\cite{lci17} is an experimental communication library aiming to study how to provide better communication support for dynamic irregular applications, among which AMTs are the most typical ones. Its topmost performance priority is to efficiently support many asynchronous communications in a multithreaded environment. It is directly built on the lowest public network interface (libibverbs~\cite{Infiniband} for Infiniband and libfabric~\cite{libfabric_cxi} for Slingshot-11) to maximize control over communication behavior. To minimize thread contention, it replaces the traditional lock-based communication runtime design with atomic-based data structures. Its flexible interface enables AMTs to implement their communication abstractions more directly and effectively, tailored to their needs. 

LCI has been integrated into two established AMT systems, HPX~\cite{Kaiser2022HPX} and PaRSEC~\cite{bosilca2013PaRSECExploitingHeterogeneitya}. These integrations demonstrate significant performance improvements~\cite{yan2023hpx_lci,mor2023PaRSEC_LCI,strack2024hpx_fft,daiss2024octotiger,yan2025hpx_lci}. For example, HPX+LCI~\cite{yan2025hpx_lci} achieved 1.5x-2.5x performance speedup in Octo-Tiger~\cite{marcello2021octo,daiss2024octotiger}, a real-world astrophysics application; PaRSEC+LCI~\cite{mor2023PaRSEC_LCI} delivered up to 12\% in HiCMA~\cite{qinglei2021hicma}, a tile-based low-rank Cholesky factorization package.

The existing LCI library was written in C with a C interface~\cite{lci17} similar to MPI and GASNet-EX. Over the years, the old interface and code base have started to restrict the exploration of a larger design space of communication libraries. In addition, the interface is not particularly friendly to external users. To address these challenges, we decided to refactor the code base into C++ and develop a new interface focused on ease of use and flexibility. 

This paper introduces the proposed new interface and outlines our plans for its C++ binding. We view the interface and the binding as two orthogonal aspects: The interface defines the functionalities provided by the communication library, while the binding specifies how these functionalities are expressed in a particular programming language.

The rest of the paper is organized as follows: 
Section~\ref{sec:interface} presents the proposed new interface in a language-neutral format. Section~\ref{sec:binding} details the C++ bindings. Section~\ref{sec:evaluation} presents preliminary evaluation results of the new C++ library. Section~\ref{sec:related_work} reviews related work, and Section~\ref{sec:conclusion} concludes the paper.

%% file: sections/interface.tex
\section{The Interface}
\label{sec:interface}

\subsection{Goal}

The new interface aims to achieve the following goals:
\begin{itemize} 
\item \emph{Customizability}: Enable users to configure communication operations with maximum adaptability, offering a wide range of tunable parameters.
\item \emph{Modularity}: Have a highly modular interface. Modules should interact exclusively through their public interfaces and support user customization and extensibility.
\item \emph{Intuitiveness}: Ensure the API design is intuitive, making most features easily deducible from the core principles.
\item \emph{Ease of Use}: Provide a straightforward starting point with basic settings, allowing users to fine-tune communication as needed incrementally.
\end{itemize}

\subsection{Resource and Operation}

The new interface consists of \emph{resources} and \emph{operations}. 
Major resources include (a) \emph{device} encapsulating the low-level network resource for communication; (b) \emph{packet pool} supporting efficient allocation and deallocation of pre-registered fixed-size internal buffers; (c) \emph{matching engine} responsible for matching send/receive; (d) \emph{completion objects} for completion notification. Major operations include (a) resource allocation, deallocation, and configuration, (b) communication posting, and (c) completion checking.

Resources can be mapped to operations independently. For example, two communication operations can use the same device but different completion objects or different devices with the same completion object. Each resource can have many tunable parameters, which we call \emph{attributes}. Users can specify the default attributes at the global scope with compilation or runtime variables and specify the attributes for each resource during its allocation.

Major communication posting operations include \emph{send}/\emph{recv}, \emph{active message}, \emph{put} (aka RDMA write), and \emph{get} (aka RDMA read). All communication posting operations are asynchronous. A local completion object is passed to the communication posting operation for subsequent completion checking.  The active message, put, and get operations can also accept a remote completion object handler. \emph{put}/\emph{get} with remote completion object specified become RDMA write/read with signal. 
The interface offers three built-in completion object types: \emph{synchronizer}, \emph{completion queue}, and \emph{function handler}. Synchronizer is similar to MPI requests but with the ability to wait for multiple completed operations before becoming ready. A communication operation can be combined with any completion object type. For example, while traditional active message primitives~\cite{von1992am} always use function handlers for remote completion, LCI's \emph{active message} operation supports remote completion objects of any type, such as \emph{completion queues}.

\emph{send} and \emph{recv} also accept a matching engine as an optional argument. \emph{send}/\emph{recv} operations using different \emph{devices} can still be matched if they use the same matching engine. The interface provides two built-in matching engine implementations: a queue-based one for in-order receives and a map-based one for better threading performance. The matching engine supports five \emph{match policies} for matching sends and receives: \emph{none}, \emph{rank-only}, \emph{tag-only}, \emph{rank-tag}, and \emph{custom}. \emph{none} means any sends and be matched with any receives. \emph{custom} means using a user-provided matching function to generate the matching key.

Operations need certain resources to perform. Resources are always passed to operations as optional arguments. There will be a default set of resources allocated by the runtime. Users only need to explicitly manage resources when they find it necessary. Users can also disable this default resource allocation to prevent resource waste.

Users can also overload certain communication resources to customize the communication semantics. For example, they can implement a completion object with an atomic counter to wait for all previously posted operations with this object by overloading the \emph{signal} method.

By default, the interface guarantees up to 16-bit tag and a 15-bit remote completion object handler for put with remote signal, constrained by the 32-bit limit of Infiniband's immediate data field. The interface supports up to 64-bit tag and 32-bit remote completion object handler for all other operations. In such cases, the runtime might choose to carry the metadata as part of the message payload at the cost of additional memory references.

The interface provides an explicit progress function, allowing users to determine when and how frequently to invoke the communication progress engine. Additionally, it supports explicit memory registration, enabling users to reuse registrations to reduce overhead when possible.

%% file: sections/binding.tex
\section{The Binding}
\label{sec:binding}

\subsection{The C++ Objectized Flexible Function}

In general, there is a trade-off between flexibility and ease of use: flexibility requires each operation to offer as many tunable parameters as possible, while ease of use requires us to maintain a reasonable number of functions with a reasonable number of arguments. In the current C interface, tunable options are specified at different levels, including global variables, communication resource configuration, function signatures, and arguments, to balance this trade-off. However, this approach is not scalable as we continue to develop new features and add more options in LCI.

In the new LCI C++ binding, we propose a new C++ programming idiom, which we call \emph{objectized flexible function}, to break this trade-off. It enables function calls with an arbitrary number of optional arguments specified in any order. It provides users with an easy-to-use default interface while allowing incremental fine-tuning of runtime behavior in any desired direction.

\begin{lstlisting}[language=C++, caption={Example of the C++ objectized flexible function idiom.}, label={lst:obj_flex_func}]
// foo(A a, B b=b0, C c=c0) -> D
struct foo_x {
    // option_t is our C++11 implementation of std::optional
    A a_; option_t<B> b_; option_t<C> c_;
    
    foo_x(A); // constructor with positional argument(s)
    foo_x&& A(A); // set the argument a
    foo_x&& B(b); // set the argument b
    foo_x&& C(c); // set the argument c
    D operator()() const; // invoke the actual function 
    D call() const;     // alias to operator()
}

// Example: use the default b and custom a1/c1.
D d = foo_x(a1).c(c1)(); 

\end{lstlisting}

The objectized flexible function idiom uses C++ classes to replace function definitions. The class has a constructor that takes all positional arguments. It has one method per argument to set the corresponding argument and the overloaded \emph{operator()} (or the \emph{call} method) to invoke the corresponding function. Listing~\ref{lst:obj_flex_func} illustrates our approach with an example.

An additional advantage of objectized flexible functions is that we can reuse a class object multiple times without passing unchanged arguments. 

\subsection{Guideline}

We use the following basic rules for the LCI new C++ interface.
\begin{itemize}
    \item The interface consists of \emph{resources} and \emph{operations}. Every resource has a set of \emph{attributes}. Every operation has a set of \emph{positional arguments}, \emph{optional arguments}, and \emph{return values}.
    \item Every resource is implemented as a C++ class. 
    This class only contains one member variable: a pointer to the real resource object. In other words, the resources are exposed to users as handles.
    The resource allocation function accepts optional arguments to set any attributes of the resource being created. The resource object has a set of methods \emph{get\_attr\_[attribute\_name]} for users to query attributes.
    \item Every LCI++ operation is implemented as an objectized flexible function with the name convention \emph{[operation\_name]\_x}. It also defines a normal C++ function with all positional arguments to simplify programming in the simple case.
    \item Fatal errors will cause the runtime to throw an exception.
\end{itemize}

%% file: sections/performance.tex
\section{Preliminary Evaluation}
\label{sec:evaluation}

We evaluate the performance of the work-in-progress LCI C++ library with a multithreaded ping-pong microbenchmark, comparing it with the LCI C library and MPI. The microbenchmark uses two nodes and one process or thread per core. Every process/thread on each node performs 100k ping-pong of 8-byte messages with a peer process/thread on the other node. We report the aggregated uni-directional message rate achieved by a single node. The LCI thread-based setup uses one LCI device per thread.

We run all experiments on Delta, a cluster at NCSA with 128 AMD CPU cores per node and 200 Gbps Slingshot-11 interconnect. We use the cluster-default Cray-MPICH 8.1.27 and libfabric 1.15.2.0. We run each configuration five times and report the mean and standard deviation.

\begin{figure}[htbp]
  \centering
  \begin{subfigure}[b]{0.48\linewidth}
    \includegraphics[width=\textwidth]{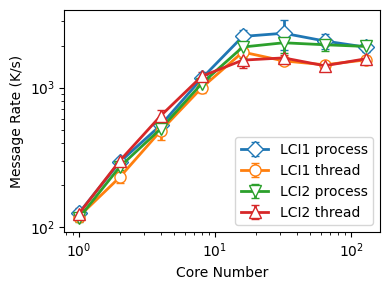}
    \caption{LCI2 v.s. LCI1}
    \label{fig:microbenchmarks-lci1}
  \end{subfigure}
  \begin{subfigure}[b]{0.48\linewidth}
    \includegraphics[width=\textwidth]{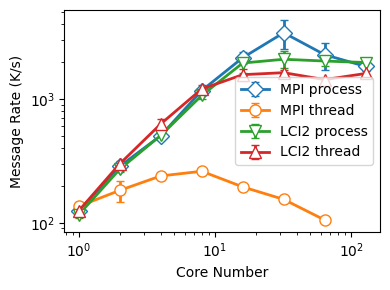}
    \caption{LCI2 v.s. Cray-MPICH}
    \label{fig:microbenchmarks-mpi}
  \end{subfigure}
  \caption{Multithreaded ping-pong results of 8B messages from 1 core to 128 cores per node. \emph{LCI2} represents the new LCI C++ library. \emph{LCI1} means the existing LCI C library. \emph{process} uses one process per core. \emph{thread} uses one thread per core.}
  \label{fig:microbenchmarks}
\end{figure}

Figure~\ref{fig:microbenchmarks} shows the results. The new LCI C++ library adds flexibility while achieving almost the same performance as the existing C library in the thread-based case. The slight performance difference in the process-based case could be further mitigated once we finish all the optimization development. In addition, a performance gap remains between the thread-based LCI and the process-based one, indicating that there may still some optimization opportunities. However, this gap has been significantly reduced compared to MPI.

%% file: sections/related_work.tex
\section{Related Work}
\label{sec:related_work}

In designing the functionalities provided by LCI, we considered many existing communication libraries. The Message Passing Interface (MPI)~\cite{mpi41} primarily offers tag-matching send-receive and collective communication as its core operations. GASNet-EX~\cite{bonachea_gasnet-ex_2018} features low-level active messages and RMA support for runtime developers. PGAS languages and libraries such as UPC~\cite{el2006upc} and OpenSHMEM~\cite{chapman2010openshmem} focus mainly on RMA with an additional abstraction layer for partitioned global memory space. 
Lower-level network portability libraries like UCX\cite{shamis2015ucx} and libfabric\cite{libfabric} provide a broader but more primitive set of point-to-point operations.
LCI aims to unify all commonly used point-to-point communication operations with many tunable options to customize communication behaviors while maintaining a user-friendly interface.

Most of the communication libraries are written in C with a C interface. The MPI specification used to define a standardized C++ binding~\cite{squyres1997MPI2_cpp} in MPI-2, but it was removed in MPI-3. There have been several proposals for MPI C++ bindings, such as Boost MPI~\cite{boost_mpi}, MPP~\cite{pellegrini2012mpp}, and MPL~\cite{ghosh2021mpl}. These bindings are largely a one-on-one mapping to the MPI C binding in an object-oriented manner, with some features to better use compilation-time optimization for certain arguments, data types, and reduction operations. Our efforts here are not limited to a new C++ binding layer for LCI but aim to refactor the entire LCI library and design a new interface in C++, mainly for flexibility and simplification purposes.

%% file: sections/conclusion.tex
\section{Conclusion}
\label{sec:conclusion}

We presented our planned next-generation C++ LCI interface. The interface is designed as a combination of resources and operations, where resources are independent and operations can flexibly combine them arbitrarily. With a default configuration and many tunable parameters, the new interface aims to achieve both flexibility and ease of use.

The proposed interface and its C++ binding are still works in progress. We will continue refining the design and welcome any feedback.